\documentclass[twocolumn,showpacs,floats,aps,amsmath,amssymb]{revtex4-1}
\usepackage{graphicx}
\usepackage{subfigure}
\usepackage{hyperref} % For hyperlinks in the PDF
\usepackage{bm}
\usepackage{float}
\usepackage{yfonts}

\begin{document}

\title{Transient dynamics and waiting time distribution of molecular junctions
in the polaronic regime}
\author{R. Seoane Souto$^1$, R. Avriller$^2$, R. C. Monreal$^1$, A. Mart\'{\i}n-Rodero$^1$ and A. Levy Yeyati$^1$}
\affiliation{$^1$Departamento de F\'{i}sica Te\'{o}rica de la Materia Condensada,\\
Condensed Matter Physics Center (IFIMAC) and Instituto Nicol\'{a}s Cabrera,
Universidad Aut\'{o}noma de Madrid E-28049 Madrid, Spain}
\affiliation{$^2$Univ. Bordeaux, LOMA, UMR 5798, F-33400 Talence, France.\\
CNRS, LOMA, UMR 5798, F-33400 Talence, France}
\date{\today}

\begin{abstract}
We develop a theoretical approach to study the transient dynamics and the time-dependent statistics for the Anderson-Holstein model 
in the regime of strong electron-phonon coupling. For this purpose we adapt a recently introduced diagrammatic approach 
to the time domain. The generating function for the time-dependent charge transfer 
probabilities is evaluated numerically by discretizing the Keldysh contour. The method allows us to
analyze the system evolution to the steady state after a sudden connection of the dot to the leads,
starting from different initial conditions. Simple analytical results are obtained in the regime of
very short times. We study in particular the apparent bistable behavior occurring for strong electron-phonon coupling,
small bias voltages and a detuned dot level. The results obtained are in remarkable good agreement with numerically
exact results obtained by Quantum Monte Carlo methods. We analyze the waiting time distribution and charge 
transfer probabilities, showing that only a single electron transfer is responsible for the rich structure found 
in the short times regime. A universal scaling (independent of the model parameters) is found for the relative amplitude 
of the higher order current cumulants in the short times regime, starting from an initially empty dot. We finally analyze the
convergence to the steady state of the differential conductance and of the differential Fano factor at the inelastic threshold,
which exhibits a peculiar oscillatory behavior.
\end{abstract}

\maketitle

\section{Introduction}
\label{sec:intro}
The study of time-dependent current 
fluctuations in nanoscale conductors 
is of great importance as it can provide information on the 
interactions and quantum correlations between electrons \cite{blanter,nazarov}. 
While these studies have been traditionally restricted to the 
stationary regime (corresponding to long measuring times),
the advent of single electron sources \cite{feve,bocquillon,dubois} has triggered
the interest in the short-times behavior.
On the one hand this knowledge would be useful to fully characterize the single
electron emitters in the high frequency range \cite{reulet}.
On the other hand, understanding the short time dynamics is a necessary requirement 
for the use of nanodevices in the detection of individual 
electrons \cite{marquardt}. 

This context suggests the need of developing new methods to characterize the statistics in the time-domain.
The concept of Waiting Time Distributions (WTD) is well known in the field of quantum optics and
stochastic processes \cite{vanKampen} but has been 
more recently introduced in electronic transport \cite{brandes}. 
Here it has been studied in the incoherent regime using master equations both within Markovian 
\cite{brandes,markus,rajabi} and non-Markovian \cite{thomas2013} approximations. 
The extension of these studies to the coherent (but non-interacting) regime is a quite recent development. 
For this case a scattering approach has been introduced \cite{albert} and adapted 
later to tight-binding models \cite{thomas2014}. Other approach to the problem is provided by 
non-equilibrium Green functions methods, which have been discussed in Ref. \cite{mukamel} and 
applied to analyze the transient dynamics of non-interacting quantum dots in Refs. \cite{tang,tang2}. 

Green function methods are in principle the most appropriate to study the effect of
interactions in the time-dependent statistical properties of
quantum coherent conductors. However, their application to this case has remained essentially unexplored. 
In the present work we provide some initial steps in this direction
by analyzing the Anderson-Holstein model in the polaronic regime.
This simple model provides the basis to understand quite complex non-equilibrium phenomena
occurring in actual systems such as phonon-assisted tunneling \cite{leroy,sapmaz} and
Franck-Condon blockade \cite{leturcq}. 

While the stationary transport properties of the Anderson-Holstein model have been extensively analyzed 
\cite{galperin1,galperin2,carmina,maier,dong,ferdinand-spectral},
their analysis in the time domain has been much less analyzed \cite{jauho}. 
Recent calculations, based on numerically exact methods like
diagrammatic quantum Monte Carlo (diagMC) \cite{ferdinand1} have indicated that for strong electron-phonon coupling there exists a
regime in which different initial conditions lead to different transport properties at short times. In a subsequent work
\cite{Ferdinand} it was demonstrated that this apparent bistability actually corresponds to a long transient
dynamics leading  to blocking-deblocking events associated to the polaron dynamics. More recent work \cite{perfetto}
has confirmed the analysis and also explored the effect of including the dot-leads Coulomb repulsion. The approach
to the steady state has also been analyzed for the Anderson-Holstein model including a continuous distribution of 
phonon modes \cite{wilner}.  

%Theoretical background on Holstein (polaronic): Muhlbacher-Rabani, Albrecht et al., Wilner et al., Wang-Thoss, Perfetto-Stefanucci. 
%The effect of electron-phonon interactions in electronic noise have been analyzed in the perturbative regime: Avriller-ALY, 
%Schmidt-Komnik, Haupt et al, Urban et al., Novotny et al.. Some of the predictions have been tested in atomic chains: Kumar et al.

In spite of these efforts, none of these works have analyzed the time evolution of the noise properties of the Anderson-Holstein
model as the system approaches the steady state. This deficit connects with the above mentioned lack of studies of
time-dependent statistics for interacting systems in the quantum transport regime. Unfortunately, numerically exact
methods like quantum Monte Carlo \cite{rabani,hutzen} or Numerical Renormalization Group (NRG) \cite{anders}
have not yet been adapted to noise studies. 

To circumvent these difficulties, the present work introduces a generalization to the time-dependent case 
of a simple analytical approach called Dressed Tunneling
Approximation (DTA), which was shown to give a good description of the
spectral and transport properties in the stationary limit for the polaronic regime \cite{seoane}.
As a first step we check that the method provides results for the time-evolution of the mean current and the dot charge
which are in good agreement with numerically exact results of Ref. \cite{ferdinand1}.  
We then study the transient WTD and the evolution of the current cumulants, showing that interactions
tend to increase the characteristic times for relaxation towards the steady state
and also enhance the asymmetry in the charge transfer probability distributions.
We also study the scaling of the relative amplitudes of the transient cumulants of higher order and find
a very robust universal behavior, that we demonstrate using analytical arguments. 
Finally, we analyze the convergence to the steady state of the differential conductance
and of the differential Fano factors at the inelastic threshold $V=\omega_0$ and also at $V=2\omega_0$.
 
\section{Model and basic theoretical formulation}
\label{sec:theory}
We consider the simplest spinless Anderson-Holstein 
model in which a single electronic level is coupled to a localized 
vibrational mode. 
Electrons can tunnel from this resonant level into a left (L)
and a right electrode (R). We shall generically refer to this central region, 
which can represent either a molecule, an atomic chain or a quantum dot, as
the ``dot" region.
The corresponding Hamiltonian is given by
$H=H_{leads}+H_{dot}+H_T$, with (in natural units, $\hbar=k_B=e=m_e=1$)
\begin{equation}
 H_{dot}=\left[\epsilon_0+\lambda\left(a^\dagger+a\right)\right]d^\dagger d+\omega_0\;a^\dagger a \;,
  \label{Hv}
\end{equation}
where $\epsilon_0$ is the bare electronic level, $\lambda$ is the
electron-phonon coupling constant and $\omega_0$ is the frequency of the 
localized vibration. The electron (phonon) creation operator in the dot
is denoted by
$d^{\dagger}$ ($a^{\dagger}$). On the other hand,
$H_{leads}=\sum_{\nu k}\epsilon_{\nu k} c_{\nu k}^\dagger c_{\nu k}$
corresponds to the non-interacting leads Hamiltonian ($\nu\equiv L,R$) where 
$\epsilon_{\nu k}$ are the leads electron energies and 
$c^{\dagger}_{\nu k}$ are the corresponding creation 
operators. The bias voltage applied to the junction is imposed by shifting 
symmetrically the 
chemical potential of the electrodes $V=\mu_L-\mu_R$.

The tunneling processes are described by
\begin{equation}
  H_T=\sum_{\nu k} \left(\gamma_{\nu k} \;c_{\nu k}^{\dagger}\;d+\mbox{h.c.}\right) \; ,
\end{equation}
where $\gamma_{\nu k}$ are the tunneling amplitudes. 

To address the polaronic regime we perform the so-called
Lang-Firsov unitary transformation \cite{LangFirsov} which eliminates
the linear term in the electron-phonon coupling \cite{Mahan}
\begin{equation}
 \bar{H}=S H S^\dagger , \quad S=e^{g d^\dagger d (a^\dagger - a)} ,\quad g=\frac{\lambda}{\omega_0} \;.
\end{equation}

Using this transformation
\begin{equation}
 \bar{H}_{dot} =\tilde{\epsilon}\;d^\dagger\;d \; + \; \omega_0 a^\dagger a 
\;,
\end{equation}
where $\tilde{\epsilon}=\epsilon_0-\lambda^2/\omega_0$. The tunneling 
Hamiltonian is transformed as
\begin{equation}
 \bar{H}_T=\sum_{\nu k} \left(\gamma_{\nu k}\;c_{\nu k}^{\dagger}\;X d+\mbox{h.c.}\right) \; ,
\end{equation}
where $X = \exp{\left[g (a - a^{\dagger})\right]}$ is the phonon cloud 
operator. On the other hand, the free leads Hamiltonian remains invariant.
For later use it is useful to introduce the tunneling rates 
$\Gamma_{\nu} 
= \mbox{Im} \sum_k |\gamma_{\nu k}|^2/(\omega - i0^+ - \epsilon_{\nu k})$
which are approximated by constants in the so-called wide band approximation, and denote $\Gamma=\Gamma_L+\Gamma_R$.

In the present work we focus on the transient dynamics which corresponds to the evolution of the system
from an initial $t=0$ state when the dot is suddenly connected to both leads.
The corresponding statistical properties of the transferred charges can be obtained from the generating function (GF)

\begin{equation}
Z(\chi,t) = \sum_{q=-\infty}^{\infty} e^{iq\chi} P_q(t) \; ,
\end{equation}
where $P_q(t)$ denotes the probability of transferring $q$ charges through the dot
in the measuring time $t$.
The GF is in turn related to the Cumulant Generating Function (CGF) by
$S(\chi,t) = \log Z(\chi,t)$, which generates the 
time-dependent charge cumulants $C_k(t) = (-i)^k \partial^k S/\partial \chi^k \rfloor_{\chi=0}$. 
We further define $\langle\langle I^k(t)\rangle\rangle = \partial C_k(t)/\partial t$, which tend to the
zero-frequency steady state current cumulants when $t\rightarrow \infty$.
Another quantity of interest to characterize the transient statistics
is the waiting time distribution $W(t)$. This can be related to the so-called idle
time probability $\Pi(t) = P_0(t)$,
defined as \cite{vanKampen}

\begin{equation}
\Pi(t) = \int_0^{2\pi} \frac{d\chi}{2\pi} Z(\chi,t)  \;.
\end{equation} 

While in a stationary situation the definition of the WTD requires a two-time measurement \cite{albert},
in the transient case the initial time is fixed at $t=0$ and one can define a single time measurement WTD
as \cite{vanKampen,tang}
\begin{equation}
W(t) = -\frac{d\Pi(t)}{dt} \;,  
\end{equation}
which gives the probability that the first electron is detected at a certain time $t$ after the connection
to the leads.

\begin{figure}[hb!]
     \begin{minipage}{1.0\linewidth}
      \includegraphics[width=1.0\textwidth]{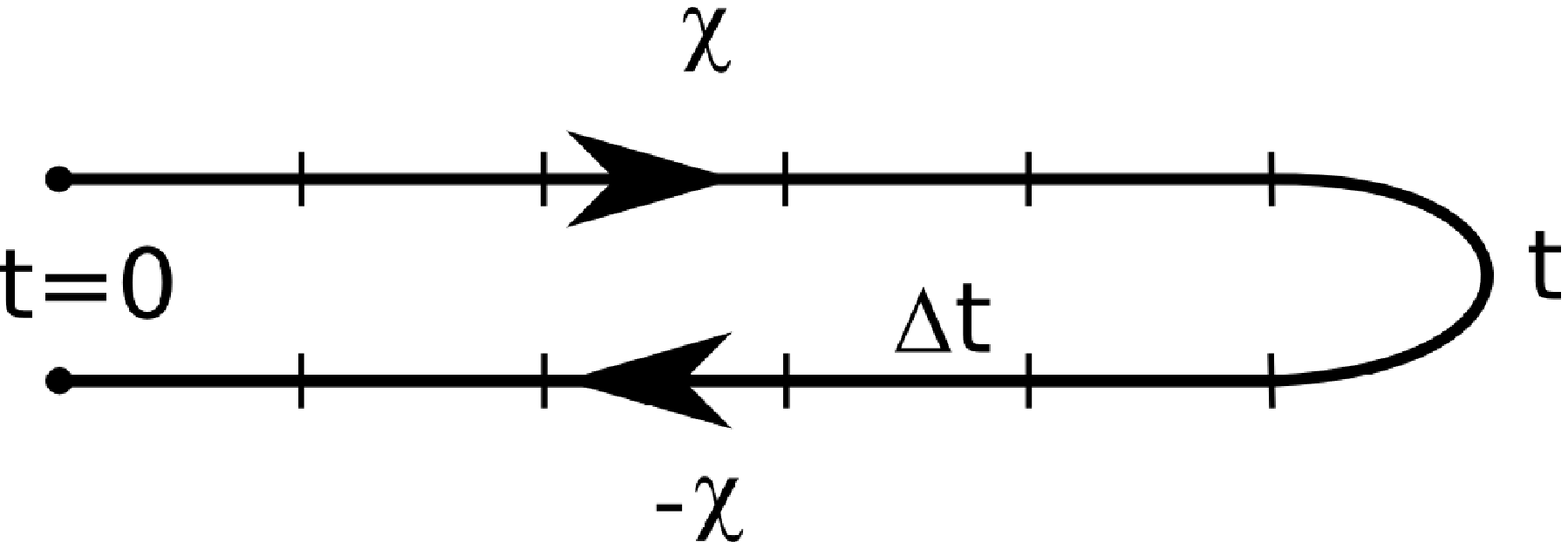}
     \end{minipage}
\caption{Keldysh contour considered to analyze the transient regime. $\chi$ indicates the counting field
changing sign on the two branches of the contour and $\Delta t$ corresponds to the time step in the
discretized calculation of the generating function $Z(\chi,t)$.}
\label{Keldysh-contour}
\end{figure}

The GF can be written \cite{mukamel} as an average of the evolution operator over the 
Keldysh contour, shown in Fig. \ref{Keldysh-contour}

\begin{equation}
Z(\chi,t) = < T_{\cal{C}} \exp \{-i\int_{\cal{C}} \bar{H}_{\chi}(t') dt'\} > \; ,
\label{z-exact}
\end{equation}
where $\bar{H}_{\chi}$ is the system Hamiltonian with a counting field $\chi(t)$
which take the values $\pm\chi$ on the two branches of the Keldysh contour entering
as a phase factor modulating the tunnel Hamiltonian, i.e.

\begin{equation}
 \bar{H}_{T,\chi} = 
\sum_{\nu k} \left(e^{i \chi_{\nu}/2} \gamma_{\nu k}\;c_{\nu k}^{\dagger}\;X d+\mbox{h.c.}\right) \; .
\end{equation}

Notice that different charge and current cumulants can be defined depending on how the phase $\chi(t)$
is distributed on the left and on the right tunnel couplings. For instance, taking $\chi_L=\chi(t)$ and $\chi_R=0$,
$Z(\chi,t)$ generates the current and charge transfer cumulants through the interface between the left lead and the dot.
This is the choice that we shall select for the rest of the paper, unless specified explicitly.

\subsection{Non-interacting case}

In Refs.\cite{utsumi,mukamel} it has been shown by path-integral methods that 
in the non-interacting case $Z(\chi,t)$ can be expressed as the following 
Fredholm determinant, defined on the Keldysh contour

\begin{equation}
Z(\chi,t) = \det\left(G \tilde{G}^{-1} \right)=
\det\left[ G\left(g^{-1}_0 - \tilde{\Sigma}\right)\right] \; ,
\label{z-nonint}
\end{equation}
where $\tilde{G}$ and $G$ denote the dot Keldysh Green functions,   
$g_0$ corresponds to the uncoupled dot case and 
$\tilde{\Sigma}$ are the self-energies due to the coupling to the leads.  
In the quantities $\tilde{G}$ and $\tilde{\Sigma}$ the $tilde$ indicates 
the inclusion of the counting field in the tunnel amplitudes. 

As shown in \cite{kamenev} a simple discretized version of the inverse free dot Green 
function on the Keldysh contour is 

\begin{equation}
i g^{-1}_0 = \left(\begin{array}{cccc|cccc} -1 & & & & & & & -\rho \\
h_- & -1 & & & & & &  \\
& h_- & -1 & & & & &  \\
& & \ddots & \ddots & & & &  \\
\hline  
&  & & 1 & -1 & & &  \\
&  & & & h_+ & -1 & &   \\
&  &  &  & & \ddots & \ddots &  \\
&  &  & & & & h_+ & -1 \end{array} \right)_{2N\times2N} \; ,
\label{kamenev}
\end{equation}
where $h_{\pm} = 1 \mp i\epsilon_0 \Delta t$, $\Delta t$ indicates
the time step in the discretization with $N=t/\Delta t$. In this
expression $\rho$ determines the initial dot charge $n_d$ by $n_d = \rho/(1+\rho)$.

On the other hand, the self-energies are given by

\begin{equation}
\tilde{\Sigma}^{\alpha\beta}(t,t') = \alpha\beta \theta(t)\theta(t') \sum_{\nu k}
\gamma_{\nu k}^2 e^{i\left(\alpha-\beta\right)\chi_{\nu}/2}g^{\alpha\beta}_{\nu k}(t,t') \; ,
\label{non-interacting-sigma}
\end{equation} 
where $g^{\alpha\beta}_{\nu k}(t,t') = -i \langle T_{\cal C} c_{\nu k}(t_{\alpha}) c^{\dagger}_{\nu k}(t'_{\beta}) \rangle$, 
with $\alpha,\beta\equiv+,-$ are the Keldysh Green functions of the uncoupled leads. 
In Appendix \ref{appendixA}
we discuss the discretization procedure and give the explicit expressions for these self-energies in the
discretized contour.  

It should be noticed that while $\tilde{\Sigma}$ are $2 \times 2$ block Toeplitz matrices depending only on 
the time arguments difference, 
$g^{-1}_0$ deviates from a perfect block Toeplitz matrix due to the $(N+1,N)$ and $(1,2N)$ entries associated with
the closing of the Keldysh contour and the initial condition respectively. The connection with the theory of 
Toeplitz determinants is an interesting issue \cite{mirlin} that goes beyond the scope of the present work and will be 
discussed elsewhere.

\subsection{Interacting case: Dressed tunneling approximation}

We now discuss the generalization of the theory to the interacting case
within the approximation introduced in in Ref.\cite{seoane}. 
For this purpose we start from the counting field functional derivative of the GF
\begin{widetext}
\begin{equation}
\frac{\delta Z}{\delta \chi} =\int_{\cal C} dt_1 \sum_k \gamma_{L k} < T_{\cal{C}}
\left(X(t_1) e^{i \chi(t_1)/2} c^{\dagger}_{Lk}(t_1) d(t_1)  -  X^{\dagger}(t_1)  
e^{-i \chi(t_1)/2} c_{Lk}(t_1) d(t_1)\right) > \; .
\end{equation}

This can be related to the three point Green function $\langle T_{\cal C} X(t) c^{\dagger}_{Lk}(t_1) d(t') \rangle$, whose equation of motion is

\begin{equation}
\left(i\partial_{t_1} - \epsilon_{Lk}\right) \langle T_{\cal C} X(t) c^{\dagger}_{Lk}(t_1) d(t') \rangle = \gamma_{Lk} e^{-i\chi(t_1)/2} 
\langle T_{\cal C} X(t) X^{\dagger}(t_1) d^{\dagger}(t_1) d(t') \rangle \;, 
\end{equation}
which can be integrated yielding
\begin{equation}
\langle T_{\cal C} X(t) c^{\dagger}_{Lk}(t_1) d(t') \rangle = \gamma_{Lk} \int_{\cal C} dt_2 
e^{-i\chi(t_1)/2}  g_{Lk}(t_2,t_1) 
\langle T_{\cal C} X(t) X^{\dagger}(t_2) d^{\dagger}(t_1) d(t') \rangle \;. 
\end{equation}
\end{widetext}

Within the decoupling procedure corresponding to the DTA
one has $\langle T_{\cal C} X(t) X^{\dagger}(t_2) d^{\dagger}(t_1) d(t') \rangle \simeq
\langle T_{\cal C} X(t) X^{\dagger}(t_2) \rangle \langle T_{\cal C} d^{\dagger}(t_1) d(t') \rangle$ which finally
allows us to write 

\begin{equation}
\frac{\partial Z}{\partial \chi} = -\int_0^t dt_1 \int_0^t dt_2 \mbox{Tr}_K 
\left\{ \frac{\partial \tilde{\Sigma}_{DTA}}{\partial \chi}(t_1,t_2) \tilde{G}(t_2,t_1)\right\} \;,
\label{z-DTA}
\end{equation}
where $\mbox{Tr}_K$ denotes trace over the $2\times2$ Keldysh space and $\tilde{\Sigma}_{DTA}$ is the DTA self-energy
whose components are given by

\begin{equation}
\tilde{\Sigma}^{\alpha\beta}_{DTA}(t,t') = \tilde{\Sigma}^{\alpha\beta}(t,t')\Lambda^{\alpha\beta}(t,t') \;,
\label{DTA-self-energy}
\end{equation} 
with $\Lambda^{\alpha\beta}(t,t') = \langle T_{\cal{C}} X(t) X^{\dagger}(t')
\rangle$ being the phonon cloud propagator. In this expression $\tilde{\Sigma}^{\alpha\beta}$ denote the self-energies 
in the non-interacting case given by Eq. (\ref{non-interacting-sigma}). On the other hand, the propagator
$\Lambda^{\alpha\beta}(t,t')$ will be evaluated assuming equilibrated phonons (see Appendix \ref{appendixB}).

Integrating Eq. (\ref{z-DTA}) and imposing the condition $Z(0,t) = 1$
one arrives to the same expression for $Z(\chi,t)$ as in Eq. (\ref{z-nonint})
but replacing $\tilde{\Sigma}$ and $G$ by $\tilde{\Sigma}_{DTA}$ and $G_{DTA}$.
All the effect of interactions are thus encoded in the DTA self-energy 
$\tilde{\Sigma}_{DTA}$. More details on the approximation are given in Appendices \ref{appendixB} and \ref{appendixC}.
It should be noted that this simple structure is valid within DTA but
in a more general approximation vertex corrections would prevent the
counting field integration leading to Eq. (\ref{z-nonint}).

\

\subsection{Tunnel and short time limits}
\label{short-time-limit}

To the lowest order in $\Gamma$ we have the expansion

\begin{widetext}
\begin{equation}
Z(\chi,t) \simeq 1 + \int_0^t dt_1 \int_0^t dt_2 \mbox{Tr}_K \left\{\left(\tilde{\Sigma}_{DTA}(t_1,t_2)-\Sigma_{DTA}(t_1,t_2) \right) g_0(t_2,t_1)\right\}
\end{equation}
which reduces to

\begin{equation}
Z(\chi,t) \simeq 1 + \int_0^t  dt_1 \int_0^t dt_2 \left(\Sigma^{+-}_{L,DTA}(t_1,t_2)g^{-+}_0(t_2,t_1)\left(e^{i\chi}-1\right)+
\Sigma^{-+}_{L,DTA}(t_1,t_2)g^{+-}_0(t_2,t_1)\left(e^{-i\chi}-1\right)\right)
\label{z-short-time}
\end{equation}

It is interesting to notice that the $\Gamma \rightarrow 0$ and the $t \rightarrow 0$ limits should coincide, i.e. 
the short time behavior is well described by the expansion to the lowest order in $\Gamma t$. This allows us
to obtain the short time limit of the GF as

\begin{eqnarray}
Z(\chi,t) &\simeq& 1 + \Big{\lbrace}
(e^{i\chi}-1) A_{L01}(t)\lbrack 1 - n_d \rbrack + (e^{-i\chi}-1) A_{L10}(t) n_d
\Big{\rbrace} \;,
\label{CGFa}
\end{eqnarray}
\end{widetext}
with
\begin{eqnarray}
\hspace{-0.2cm}
A_{L01}(t) &=& \frac{2\Gamma_L}{\pi} \times \nonumber \\
&& \sum_{n=0}^{\infty}\alpha_n \int_{-W}^{W} d\omega \frac{1-\cos\lbrack
(\omega-\tilde{\epsilon}-n\omega_0)t\rbrack}{(\omega-\tilde{\epsilon}-n\omega_0)^2}
f_{L}(\omega)
\label{CGFb} \nonumber \\
A_{L10}(t) &=& 
\frac{2\Gamma_L}{\pi} \times \nonumber\\
&& \hspace{-1cm} \sum_{n=0}^{\infty}\alpha_n \int_{-W}^{W} d\omega \frac{1-\cos\lbrack
(\omega-\tilde{\epsilon}+n\omega_0)t\rbrack}{(\omega-\tilde{\epsilon}+n\omega_0)^2}
\left[f_{L}(\omega)-1\right] \;, \nonumber
\label{CGFc}
\end{eqnarray}
where $f_L(\omega)$ is the Fermi distribution at the left electrode and, at zero temperature, 
$\alpha_n = e^{-g^2} g^{2n}/n!$.
The physical interpretation of the amplitudes $A_{L01}$ and $A_{L10}$ is transparent in the $t\rightarrow\infty$ limit
where $A_{L01}/t$ and $A_{L01}/t$ tend
to the Fermi Golden rule rates derived for the sequential tunneling
regime (see e.g. Ref.\cite{koch}).

As expected in the short time limit $\Gamma t \ll 1$, the GF in Eq.(\ref{CGFa})  involves charge transfer of a single electron,
with only $P_0(t)$, $P_{-1}(t)$ and $P_{1}(t)$ having a significant weight, and are given by 

%\begin{widetext}
\begin{eqnarray}
P_q(t) &=&   A_{L01}(t)\lbrack 1 - n_d \rbrack \delta_{q,1} 
- A_{L10}(t) n_d \delta_{q,-1}  \; ,
\label{CGFf}
\end{eqnarray}
%\end{widetext}
and $P_0(t) = 1 - P_1(t) - P_{-1}(t)$. 
Correspondingly, the WTD is proportional 
to the left current $I_L(t)$, i.e. 

\begin{eqnarray}
W(t) &=& -\frac{d}{dt} P_{0}(t) = |\frac{I_L(t)}{e}|.
\label{CGFg}
\end{eqnarray}

At zero temperature and neglecting contributions from the band edges
(which is justified in the wide band approximation) we obtain

\begin{widetext}
\begin{equation}
I_L = \Gamma_L (1-2n_d) + \frac{2\Gamma_L}{\pi} \sum_n \alpha_n 
\left\{\mbox{Si}\left[(\mu_L+n\omega_0-\tilde{\epsilon})t\right](1- n_d) 
- \mbox{Si}\left[(\mu_L-n\omega_0-\tilde{\epsilon})t\right] n_d\right\} 
\label{tunnel-limit}
\end{equation} 
\end{widetext}
where Si denotes the sine integral function.

Due to the property $\lim_{x \rightarrow 0} \mbox{Si}(x) = 0$ we
find that the initial current is $I_L = \pm \Gamma_L$, with the sign depending on the initial charge
($n_d=0$ or $n_d=1$). This property
fixes also the initial value of the WTD as $W(t) \equiv \mbox{abs}\left(I_L(0)\right)$. 
It should be noticed that we are
neglecting the system evolution on time scales smaller than the inverse of the leads bandwidth (see Appendix \ref{appendixA}),
which explains why the initial current can be non-zero. However, for the symmetrized current 
$I(t)=(\Gamma_R I_L(t)-\Gamma_L I_R(t))/\Gamma$ the initial value is zero and the first charge transfer cumulant is
a continuous function starting from zero regardless of the initial condition.

\section{Results}
\label{sec:res}

\subsection{Evolution of mean current and charge: comparison to diagMC results}

\begin{figure}[h!]
     \begin{minipage}{1.0\linewidth}
      \includegraphics[width=1.0\textwidth]{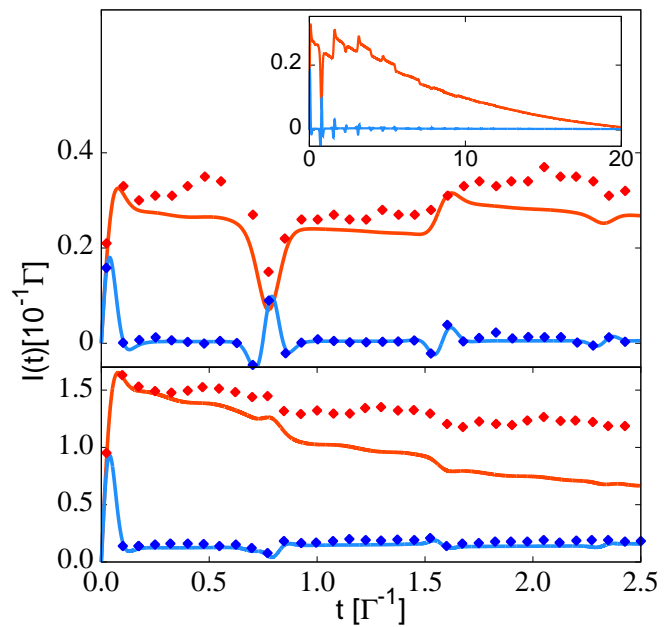}
     \end{minipage}
\caption{(Color online) 
Symmetrized transient current $I(t)$: comparison between DTA (full lines)
and diagMC (dots) for initially
empty (red) and initially full (blue) dot. $\tilde{\epsilon}=-10\Gamma$, $g=2$,
$\omega_0=8\Gamma$ and 
$V=5\Gamma$ (upper panel) 
and $V=26\Gamma$ (lower panel). The inset illustrates the convergence to the
steady state for the case $V=5\Gamma$.} 
\label{transient-current}
\end{figure}

\begin{figure}[h!]
     \begin{minipage}{1.0\linewidth}
      \includegraphics[width=1.0\textwidth]{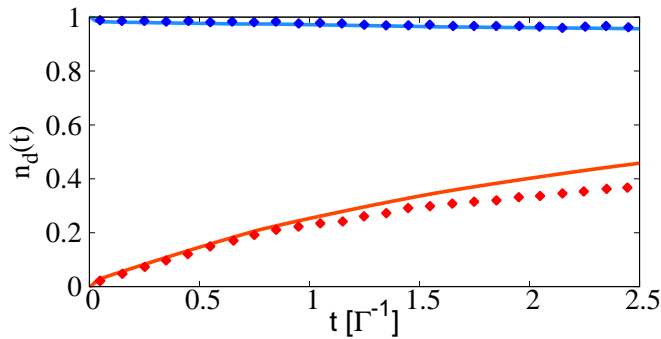}
     \end{minipage}
\caption{(Color online) 
Time-dependent dot occupation: comparison between DTA and diagMC for initially
empty and initially full dot. Same parameters as in Fig. \ref{transient-current} (lower panel).}
\label{transient-charge}
\end{figure}

We start by analyzing the transient behavior of the Holstein model
for different initial conditions. In Ref. \cite{ferdinand1} 
it was shown that for
certain parameters range the model exhibits an apparent bistable behavior.
Further analysis \cite{Ferdinand,perfetto}
revealed that this apparent bistability was caused by a
long transient regime associated with the slow polaronic dynamics. 
The comparison of the symmetrized current and the charge evolution obtained within DTA
with the numerically exact results obtained by diagMC are shown in Figs.
\ref{transient-current} and \ref{transient-charge}.
These results corresponds to the case of a deep level $\tilde{\epsilon}=-10\Gamma$ 
and large phonon frequency $\omega_0 = 8\Gamma$ where
the apparent bistable behavior is more pronounced. There is a remarkable
agreement between the DTA and the diagMC results for the current
in the $V=5\Gamma$ case. The agreement with the current
is somewhat poorer for $V=25\Gamma$, which reflects a limitation of the DTA
for describing the spectral density around the Fermi energy in this
regime (see \cite{seoane}) but there is a very good agreement in the
evolution of the mean charge for this case
(see Fig. \ref{transient-charge}). This
represents an improvement with respect to the approximation used in
Ref. \cite{Ferdinand}.

The apparent bistable behavior at short times corresponds actually to
a long transient dynamics, as illustrated by the DTA results on a longer time
scale (see inset in Fig. \ref{transient-current}).

\subsection{Waiting time distribution and transient statistics}

\begin{figure}[h!]
     \begin{minipage}{1.0\linewidth}
      \includegraphics[width=1.0\textwidth]{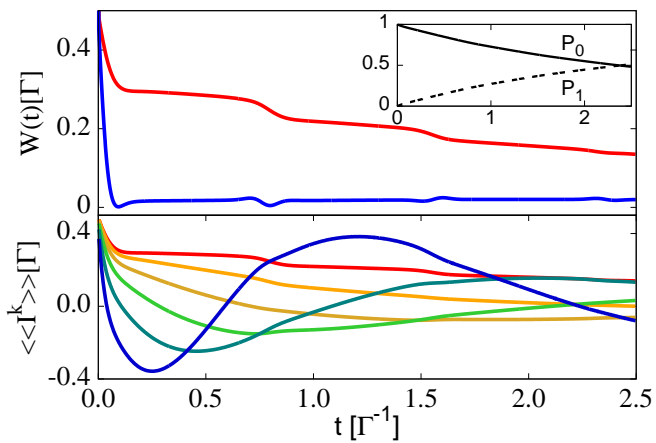}
     \end{minipage}
\caption{(Color online) 
Upper panel: waiting time distribution for the initially empty (red) 
and initially occupied (blue) cases for the same parameters as in Fig.
\ref{transient-current} (lower panel). Inset: probabilities $P_0$ (full line) and
$P_1$ (dashed line) for the initially empty state.
The lower panel shows the transient dynamics of current cumulants
$\langle\langle I^k(t) \rangle\rangle$ with $k=1, 2, 3, 4, 5, 6$ (from top to bottom at short times) for the initially empty case.} 
\label{Imoments-ferdinand}
\end{figure}

Further insight on the transient dynamics is provided by analyzing
the evolution of the WTD and the higher current cumulants, see 
Fig. \ref{Imoments-ferdinand}. 

As can be observed, in the initially empty case the WTD exhibits 
a non-monotonous decrease with small steps at time $\sim n 2\pi/\omega_0$
associated with the polaron dynamics. The current cumulants in this short
time regime have an increasing amplitude with increasing cumulant order (we come back to this point
below). The inset in the upper panel of Fig. \ref{Imoments-ferdinand}
indicates that this short time dynamics is associated with a
single electron transfer, with only $P_0$ and $P_1$ being non-negligible. Thus
the waiting time distribution follows essentially the current at short times. On the other
hand, for the initially occupied state the current dynamics is almost
blocked in this short time regime.

\begin{figure}[h!]
     \begin{minipage}{1.0\linewidth}
      \includegraphics[width=1.0\textwidth]{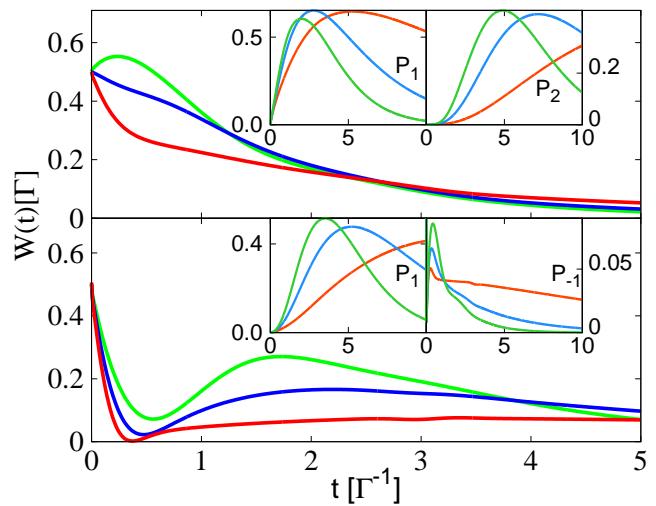}
     \end{minipage}
\caption{(Color online) 
Waiting time distribution for increasing $g$ (0 (green), 1 (blue)
and 1.5 (red))
for initially empty (upper panel)
and initially full (lower panel) with  
$\tilde{\epsilon}=-\Gamma$, $\omega_0=2\Gamma$ and $V=5\Gamma$.
The insets show the corresponding probabilities $P_1(t)$ and $P_{2}(t)$ 
(upper panel) and $P_1(t)$ and $P_{-1}(t)$ (lower panel).}
\label{waiting_g}.
\end{figure}

All the previous results indicate that the typical times for the transient
regime are increased by the effect of interactions. A more clear picture
of this effect is provided by Fig. \ref{waiting_g} where $W(t)$
is shown for increasing values of $g$ for the initially empty and
initially occupied states. 

At short time scales ($t\le \Gamma^{-1}$) and for weak coupling to the leads, in the initially empty case,
the WTD is well approximated by Eq. (\ref{tunnel-limit}). 
Thus, it exhibits an initial linear behavior fixed by 
$W(t)\approx \Gamma_L/2 + 2\Gamma_L/\pi\sum_n \alpha_n (\mu_L-\tilde{\epsilon}-n\omega_0)t$ followed by 
an extremum at 
$t \approx \pi \lbrace |\sum_n \alpha_n (\mu_L-\tilde{\epsilon}-n\omega_0 )/
\sum_n \alpha_n (\mu_L-\tilde{\epsilon}-n\omega_0 )^3| \rbrace^{1/2}$.
In the non-interacting case $g=0$ only the $n=0$ term contributes and 
the WTD exhibits an initial positive slope for the parameters in 
Fig. \ref{waiting_g}, reaching a maximum peak located at 
$t \approx \pi/|V/2-\tilde{\epsilon}|$.
Increasing the coupling strength, the initial slope decreases and becomes 
eventually negative. For $g\approx 1.3$, the peak
becomes a dip indicating the transition into the strong coupling regime.
The probabilities $P_1(t)$ and $P_2(t)$ shown
in the insets allow to visualize the injection of the first and second
electron, and their ralentization with increasing interaction.

On the other hand, the initially occupied case exhibits a different short 
time linear scaling
for the WTD $W(t)\approx \Gamma_L/2 - 2\Gamma_L/\pi\sum_n \alpha_n (V/2-\tilde{\epsilon}+n\omega_0)t$
followed by a dip at very short times
$t \approx \pi \lbrace |\sum_n \alpha_n (\mu_L-\tilde{\epsilon}+n\omega_0 )/
\sum_n \alpha_n (\mu_L-\tilde{\epsilon}+n\omega_0 )^3| \rbrace^{1/2}$
which is associated to the blocking effect of the occupied level.
Contrary to the empty case, the evolution of the dip is monotonous with the
interaction strength: the dip depth increases and its position shifts 
towards smaller times with increasing $g$. As shown by the insets, the 
backward flow probability $P_{-1}(t)$ is
quite significant in this case, contrary to the initially empty case.

\begin{figure}[h!]
     \begin{minipage}{1.0\linewidth}
      \includegraphics[width=1.0\textwidth]{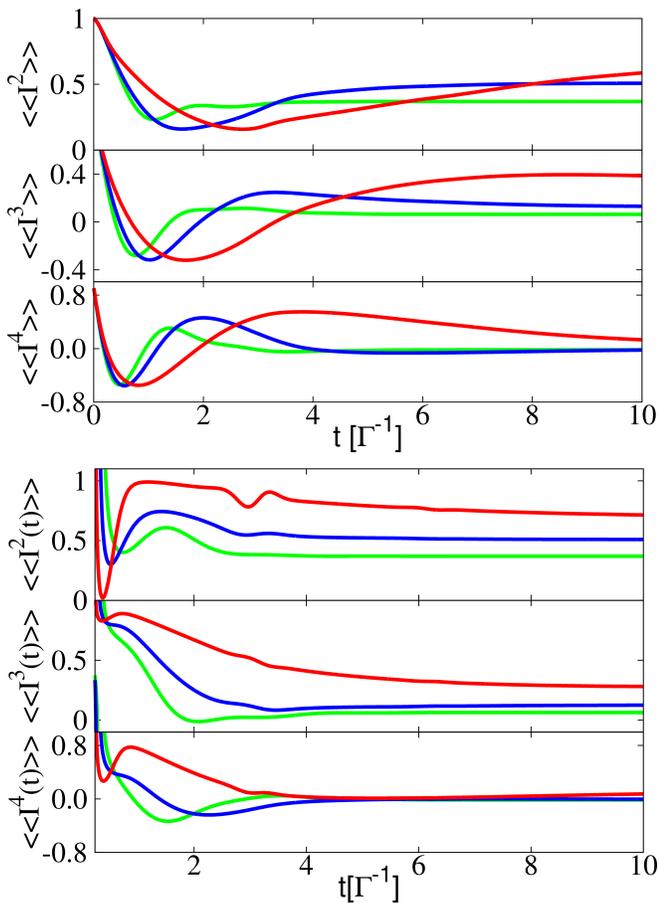}
     \end{minipage}
\caption{(Color online) 
Higher order cumulants (normalized to the
mean current) for increasing $g$ values (same color code as in Fig.
\ref{waiting_g}) for the initially
empty (upper panels) and initially full cases (lower panels). Notice that in the lower panels the time
axis starts at times slightly larger than 0 in order to avoid the divergent behavior of the normalized
cumulants in the $t\rightarrow 0$ limit.}
\label{gpanels}
\end{figure}

Similarly, the increase of $g$ has an impact in the evolution and the
stationary limit of the
higher order cumulants $\langle\langle I^k\rangle\rangle$. These are shown in Fig. \ref{gpanels} for $k=2,$ 3 and 4. 
The cumulants are normalized
to the time dependent mean current, which allows to appreciate more clearly
the differences with increasing $g$.
As a general feature, both the interacting and non-interacting cases 
exhibit an increase in the transient amplitude with increasing cumulant order, 
as already mentioned in connection to Fig. \ref{Imoments-ferdinand}.
The effect of increasing $g$ is twofold: first, it slows down the
dynamics and second, the relative asymptotic values of the cumulants are larger
than the non-interacting ones.
In the initially occupied case (lower panels in Fig.\ref{gpanels}) the same
effects can be observed. The divergent relative amplitudes at very short times 
are due to the change of sign of the mean current occurring in this case.

\subsection{Universal scaling of normalized transient cumulants}

\begin{figure}[h!]
     \begin{minipage}{1.0\linewidth}
      \includegraphics[width=1.0\textwidth]{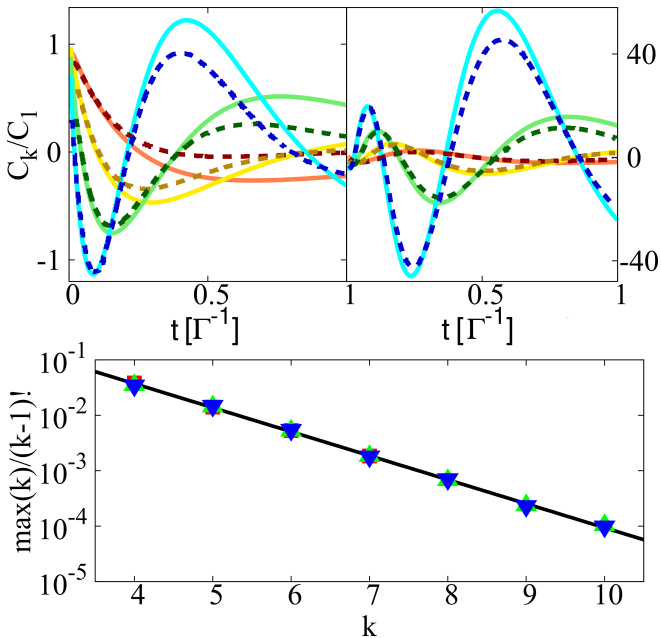}
     \end{minipage}
\caption{(Color online) Upper panels: normalized transient current cumulants obtained from Eq. (\ref{cumulants-formula}) (full lines)
and the corresponding results adapted from Ref. \cite{flindt-exp} (dashed lines) for $k=4,5,6,7$ (left panel) and $k=8,9,10,11$ (right panel). For the comparison we have assumed that $C_1(t)$ follows a simple law $A(1-e^{-\Gamma t})$,
which sets the time scale.
Lower panel: scaling of the maximum amplitude of the relative cumulants $C_k/C_1$ (indicated by $\mbox{max}(k)$ in the figure)
as a function of $k$. 
The relative cumulants are normalized with $(k-1)!$ and the linear slope in the logarithmic scale indicates
scaling as $(k-1)! \pi^{-k}$ (see text).
The different symbols correspond to the cases $g=0$, 1 and 1.5 in Fig. \ref{waiting_g}.}
\label{scaling}
\end{figure}

In spite of the renormalization of the characteristic times and of the
asymptotic cumulants introduced by interactions, for the initially empty
case some ``universal" features can be identified. For instance, as shown in the lower panel of Fig. \ref{scaling}, 
the maximum amplitude of the normalized transient cumulants $C_k(t)/C_1(t)$ 
are found to follow a scaling which is
independent of the value of the interaction parameter $g$. This scaling is also found to be extremely robust
with other model parameters like $\tilde{\epsilon}$ or $\omega_0$. 

We can explain this behavior by considering the short
time limit discussed in Sect. \ref{short-time-limit}. For the initially empty case $g^{+-}_0(t,t') =0$ and as
Eq. (\ref{z-short-time}) indicates, the current flow is unidirectional. Moreover, in this limit all derivatives 
of the GF are equal, i.e. $(-i)^n \partial^n Z(t,\chi)/\partial \chi^n\rfloor_0\equiv x(t)$, corresponding to the situation
where just a single electron is involved, as shown in the inset of Fig. \ref{Imoments-ferdinand}.
Correspondingly, the CGF at short times can be written as
\begin{equation}
S(\chi,t) \simeq \log \left[1 + x(t)\left(e^{i\chi}-1\right)\right] \;.
\end{equation}

A simple Taylor expansion allows to identify the charge cumulants as
\begin{equation}
C_k(t) = -\sum_{q=1}^{\infty} \left[-u(t)\right]^q q^{k-1},
\end{equation}
where $u(t)=x(t)/(1-x(t))$. Taking the continuous limit and evaluating the above summation as an integral (valid at sufficiently
large $k$) we find
\begin{equation}
C_k = \frac{2(k-1)!\cos\left[k\arctan\left(\frac{\pi}{\log u}\right)\right]\left(\sigma_u\right)^{k+1}}{\left(\log^2 u + \pi^2\right)^{k/2}} \;,
\label{cumulants-formula}
\end{equation}
where $\sigma_u = \mbox{sign}(\log u)$.

While the cosine factor in this expression is responsible for the oscillatory behavior of the cumulants at short times, their
maximum amplitude is controlled by the $(k-1)!$ factor and the power law in the denominator. As in the region of maximum amplitude
typically $\log^2 u \ll \pi^2$, it is straightforward to show from Eq. (\ref{cumulants-formula}) that the cumulants maximum amplitude
scale as $(k-1)! \pi^{-k}$. This law describes with accuracy the scaling of the relative current cumulants
already shown in the lower inset of Fig. \ref{scaling}. 

The factorial increase of the transient cumulants has been already pointed out and demonstrated experimentally in Ref. \cite{flindt-exp}.
In contrast to the present study, they considered the sequential tunneling regime. However, as we show in the upper panels of Fig. 
\ref{scaling}, the short time behavior of the transient cumulants predicted by Eq. (\ref{cumulants-formula}) 
remarkably agrees with
the results of Ref. \cite{flindt-exp}. The reason for this agreement is the universality of the generating function in the 
short time regime for unidirectional transport corresponding to the initially empty dot case. At longer times the results from 
Ref. \cite{flindt-exp} deviate from the predicted behavior by Eq. (\ref{cumulants-formula}) by a global extra exponential decay,
which can be associated with the differences with the setup considered in Ref. \cite{flindt-exp}. 

%Correspondingly the CGF 
%normalized cumulants satisfy the recurrence relation \cite{hald}
%\begin{equation}
%C_n(t)/C_1(t) = 1 - \sum_{j=1}^{n-1} \left(\begin{array}{c} n-1 \\ j-1 \end{array}\right) C_j(t)
%\label{recurrence}
%\end{equation}
%with the initial condition $C_1(t)=M_1(t)=\int^t_0 dt' I(t')$. 

%Due to the approximate linear increase of the current in the short time limit, 
%the recurrence relation also holds for the maxima of the relative current cumulants 
%$\langle\langle I^n(t)\rangle\rangle/\langle\langle I(t)\rangle\rangle$. 
%The short time behavior of the current cumulants thus follow from the behavior of the current itself. 
%The universal scaling of the current cumulants can be verified from the recurrence relation (\ref{recurrence}) assuming any simple 
%monotonously increasing form for $I(t)$ at short times. 
%Notice that this scaling is a single particle property not depending on the model parameters. It also holds
%in the incoherent regime analyzed in Ref. \cite{flindt-exp}.

\subsection{Conductance and Fano factor dynamics at $V=n\omega_0$}

\begin{figure}[h!]
     \begin{minipage}{1.0\linewidth}
      \includegraphics[width=1.0\textwidth]{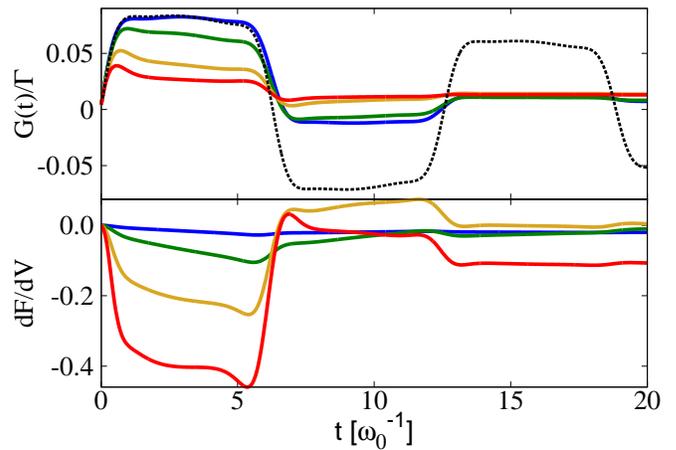}
     \end{minipage}
\caption{(Color online) Transient conductance (upper panel) and differential
Fano factor (lower panel) for $V=\omega_0$, $g=1.5$, $\tilde{\epsilon}=0$ 
and different values of $\Gamma/\omega_0$ (0.05 (blue), 0.25 (green), 1 (yellow) and 2 (red)).
The dashed line in the upper panel corresponds to the tunnel limit
analytical result of Eq. (\ref{tunnel-limit}).}
\label{w0}
\end{figure}

\begin{figure}
     \begin{minipage}{1.0\linewidth}
      \includegraphics[width=1.0\textwidth]{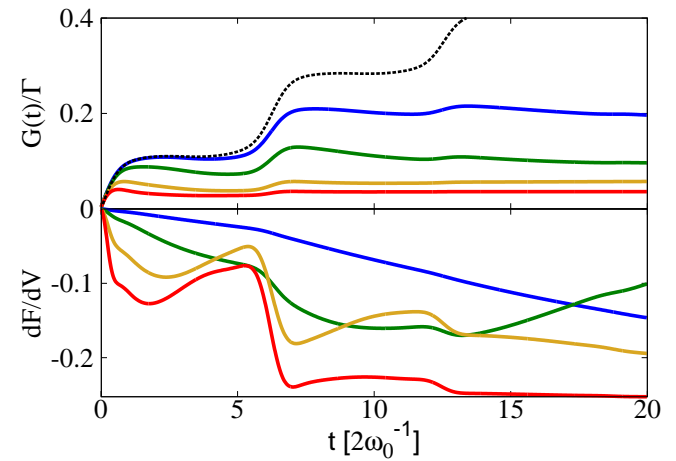}
     \end{minipage}
\caption{(Color online) Same as Fig. \ref{w0} for $V=2\omega_0$.}
\label{2w0}
\end{figure}

It is also worth analyzing the current and noise dynamics for bias voltages
close to the conditions $V \sim n \omega_0$. The behavior of the stationary
noise at the inelastic threshold $V = \omega_0$ has been analyzed in several works
in the limit of weak interaction \cite{avriller2009,komnik,belzig,kumar}, showing that it can either 
exhibit an increase or a decrease due to the opening of the inelastic channel.
In contrast, for $V=2\omega_0$, the analysis of stationary noise in the
polaronic regime presented in Ref. \cite{seoane} indicates that it exhibits 
a suppression associated to the opening of a side band (new elastic
channel).
The transient conductance for $V \sim \omega_0$ and $V \sim 2\omega_0$ 
are shown in Figs. \ref{w0} and \ref{2w0} for $\tilde{\epsilon}=0$, $g=1.5$ and
different values of $\Gamma$. For comparison we show
the prediction for the tunnel limit, given by Eq. (\ref{tunnel-limit}) as dashed lines. 
To illustrate the behavior of the noise we choose to represent the differential Fano factor
$\partial F(t)/\partial V$, where $F(t) = \langle\langle I^2\rangle\rangle/(2\langle\langle I\rangle\rangle)$, 
which allows to see more clearly 
the differences in the $\Gamma \rightarrow 0$ limit.

As can be observed in these plots, the behavior of the transient conductance 
for $\Gamma \rightarrow 0$
is remarkably different in the two cases: while for $V=\omega_0$ the conductance
exhibits a sequence of up and down steps at $t\sim 2n\pi/\omega_0$; for $V=2\omega_0$ 
the sequence corresponds to steps up only. The approximate behavior is well captured
by the analytical expression of Eq. (\ref{tunnel-limit}) shown as dashed lines
in Figs. \ref{w0} and \ref{2w0}. When $\Gamma$ increases the step structure in the conductance
is progressively damped. On the other hand, the noise exhibits an interesting
different evolution with increasing $\Gamma$. While for $\Gamma \rightarrow 0$ the
differential noise and the conductance are approximately equal (as expected for the
tunnel limit) for larger $\Gamma$ the differential Fano factor converge to either positive
or negative values for $V=\omega_0$, or systematically to negative values for $V=2\omega_0$.
This is consistent with the predictions of Ref. \cite{seoane} for the stationary case. It is
interesting to remark the difference between the present calculations and those of Ref. \cite{koch}.
In that work a sequential tunneling approach was used, including the effect of the induced nonequilibrium
phonon population leading to giant Fano factors associated 
to the Frank-Condon blockade effect. 

\section{Conclusions and outlook}
\label{conclusions}
In this work we have presented an analysis of the time-dependent statistics of electron transport
through a resonant level coupled to a localized vibrational mode. We have restricted our analysis
to the transient regime which is established after a sudden connection of the level to the leads.
For this analysis we have adapted the recently developed DTA decoupling scheme \cite{seoane}, which provides
a good description of the stationary transport properties in the polaronic regime. 
In spite of its approximate character, our analysis has revealed several features
of general validity. In the first place it shows that interactions tend to increase exponentially
the relevant time scales for the transient dynamics, leading to an apparent bistability at short
times in certain parameters regime, in agreement with a previous analysis by some of us \cite{Ferdinand}. 
Second, we have demonstrated that in the short time scale
and for an initially empty dot the higher order cumulants exhibit oscillations with a universal
scaling amplitude. This universal character arises due to the fact that a single electron transfer controls
the transport properties at these short time scales. Our analysis has furthermore revealed a 
peculiar oscillatory convergence of the conductance and the Fano factor at the inelastic
threshold $V=\omega_0$.  

The present work constitutes a first step in the study of time-dependent statistics for the
interacting quantum coherent transport regime. We envisage several possible extensions of 
this work, like the study of waiting time distributions in the stationary regime for interacting systems;
analyzing the effect of unequilibrated phonons and extending the study to superconducting systems.

\begin{acknowledgments}
R.S., R.C.M, A.M.R. and A.L.Y. acknowledge funding from Spanish MINECO through grants FIS2011-26516 and FIS2014-55486-P. R.A. acknowledges funding from French ANR grant ORGAVOLT and Partenariats Hubert Curien NANO ESPAGNE Project N0. 31404NA.
\end{acknowledgments}

\appendix
\section{Time discretization procedure}
\label{appendixA}

We describe in this appendix the time discretization procedure. For simplicity we discuss here first the 
non-interacting case. For describing the
leads we consider the simplest case of a flat density of states (wide band approximation) within
an energy range in the interval $(-W,W)$. The leads bandwidth $W$ is assumed to be much larger than 
the tunneling rates $\Gamma_{\nu}$. The discretized time-dependent self-energies correspond to the 
Fourier transform of these energy dependent self-energies evaluated at the discrete time mesh 
defined by $t_j = \Delta t j$, which at zero-temperature yields 

\begin{eqnarray}
\tilde{\Sigma}^{+-}_{ij} &=& -\sum_{\nu\equiv L,R}
\frac{\Delta t \Gamma_{\nu}e^{i\chi_{\nu}}}{\pi} \frac{e^{-i\mu_{\nu}\Delta t(i-j)} - e^{-iW\Delta t(i-j)}}{(i-j)} 
\nonumber\\
\tilde{\Sigma}^{-+}_{ij} &=& -\sum_{\nu\equiv L,R}
\frac{\Delta t \Gamma_{\nu}e^{-i\chi_{\nu}}}{\pi} \frac{e^{-iW\Delta t(i-j)} - e^{-i\mu_{\nu}\Delta t(i-j)}}{(i-j)} \;, \nonumber\\ 
\label{wideband-sigma}
\end{eqnarray}
where $i,j\ge 1$.
The other self-energy components are given by $\Sigma^{++}_{ij} = -\theta\left[i-j\right]\Sigma^{-+}_{ij} - 
\theta\left[j-i\right]\Sigma^{+-}_{ij}$ and
$\Sigma^{--}_{ij} = -\theta\left[i-j\right]\Sigma^{+-}_{ij} - \theta\left[j-i\right]\Sigma^{-+}_{ij}$, where $\theta\left[n\right]$
is the discrete form of the Heaviside function.
These last expressions are not unambiguously defined for $i=j$. We have found that the more stable algorithm corresponds to the
choice $\theta\left[0\right]=1/2$.

The finite flat bandwidth model at zero temperature produces current cumulants increasing linearly from zero, followed by an oscillatory
pattern on the time scale $1/W$, produced by the exponential term in Eq. (\ref{wideband-sigma}). 
This pattern is absent in the symmetrized current cumulants. In order to resolve these features, the 
time step has to be smaller than this typical time, which leads to the condition of $\Delta t\lesssim 1/W$ for a stable algorithm.
In order to avoid features associated with the finite bandwidth it is convenient to generalize the calculation to
finite temperatures, performing an expansion of the self-energies in Matsubara frequencies. In this case, 
the limit $W\to\infty$ is well defined, avoiding 
divergences at time equal to zero. The expressions for the self-energies at finite temperature are
\begin{eqnarray}
 \tilde{\Sigma}^{+-}_{jk}&=&2 i\sum_{\nu=L,R} e^{i\chi_\nu} \Gamma_\nu f^{\nu}_{jk}\nonumber\\
 \tilde{\Sigma}^{-+}_{jk}&=&2 i\sum_{\nu=L,R} e^{-i\chi_\nu} \Gamma_\nu(f^{\nu}_{jk}-\delta[j-k]) \;,
\end{eqnarray}
where the Fermi function can be computed as
\begin{eqnarray}
 f^{\nu}_{jk}&=&i\sum_{n=0}^{\infty}R_n\left[\theta[j-k]e^{\beta_n (j-k)\Delta t}\right.\nonumber\\
&&\left.-\theta[k-j]e^{-\beta_n (j-k)\Delta t}\right]e^{-i\mu_x (j-k)\Delta t}+\frac{\delta[j-k]}{2}. \nonumber\\ 
\end{eqnarray}
Here $\beta_n$ and $R_n$ represent the poles and the residues of the Matsubara expansion. The convergence speed can be improved by using 
the approximated poles and residues proposed by 
T. Ozaki \cite{Ozaki} and computed using a continued fraction. Differently from the zero-temperature finite bandwidth case, 
the fast oscillatory term is absent, and the maximum of the 
time step is controlled by the $\Gamma$ parameter. We have found that $\Delta t\lesssim 1/(10\Gamma)$ is 
sufficient to warrant the convergence of the numerical algorithm.\\

In the interacting case, the DTA self-energy, given by Eq. (\ref{DTA-self-energy}), can be written as an infinite sum over 
polaronic weights (see Appendix \ref{appendixB}).
The number of terms needed to be added for convergence is higher when increasing the electron-phonon coupling, 
meaning that higher sidebands become more important. Considering that the series can be truncated with precision enough at 
a given integer value, $n$, the condition to converge the calculation can be obtained following the same reasoning done in 
the non-interacting case, i.e. the polaronic terms generate oscillations of a minimum period $1/(n\omega_0)$, which have to 
be resolved for convergence, leading to the condition $\Delta t\lesssim 1/(n\omega_0)$.

\section{Dressed Tunneling Approximation}
\label{appendixB}

In this appendix we summarize the main expressions of the Dressed Tunneling Approximation (DTA) used in this work. The DTA approximation is built from the Dyson 
equation, where the leads self-energies have been dressed with the phonon cloud \cite{seoane}
\begin{equation}
 \hat{G}=\hat{g}_0+\hat{g}_0 \hat{\Sigma}_{DTA}\hat{G},
\end{equation}
where $\hat{g}_0$ corresponds to the undressed dot, $\hat{\Sigma}_{DTA}$ is the DTA self-energy (as defined in Eq. (\ref{DTA-self-energy}))
and the $\hat{\;}$ is used to denote Keldysh structure. In this expression integration over internal times is implicitly assumed.
%In this approximation, all the polaronic 
%effects are introduced through this self, which is computed as
%\begin{equation}
% \Sigma_{DTA}^{\alpha\beta}(t,t')=\alpha\beta\sum_{x=L,R}\gamma^{2}_{x}\Lambda^{\alpha\beta}(t,t')g_{x}^{\alpha\beta}(t,t'),
%\end{equation}
%where $\alpha,\beta=+,-$ are the Keldysh contour indexes, $\Lambda^{\alpha\beta}(t,t')=\left\langle T_C X(t)X^{\dagger}(t')\right\rangle$ is the polaronic correlator and $g_{L(R)}$ are the lead Green 
%functions of the left (right) electrode, which are computed in the so-called wide band approximation. 
Finally, dressing again the full Green function, we find the final expression as
\begin{equation}
 \hat{G}^{\alpha\beta}_{DTA}(t,t')=\hat{G}^{\alpha\beta}(t,t')\Lambda^{\alpha\beta}(t,t'),
\end{equation}
where the phonon cloud propagator $\Lambda^{\alpha\beta}(t,t')$ is evaluated assuming equilibrated phonons, i.e. 

\begin{equation}
\Lambda^{+-}(t,t') = \left(\Lambda^{-+}(t,t')\right)^* = \sum_{n=-\infty}^{\infty} \alpha_n e^{in\omega_0(t-t')},
\end{equation}
with 

\begin{equation}
\alpha_n = e^{-g^2\left(2n_p+1\right)} I_n\left(2g^2\sqrt{n_p(1+n_p)}\right) e^{n\beta\omega_0/2},
\end{equation}
$I_n$ being the modified Bessel function of the first kind, which is symmetric in the
$k$ argument ($I_n=I_{-n}$) and $n_p$ is the Bose factor $1/\left(e^{\beta\omega_0} -1\right)$ with $\beta=1/T$.
The remaining components $\Lambda^{++}(t,t')$ and $\Lambda^{--}(t,t')$ are determined
by $\Lambda^{++}(t,t') = \theta(t-t') \Lambda^{-+}(t,t') + \theta(t'-t) \Lambda^{+-}(t,t')$ and
$\Lambda^{--}(t,t') = \theta(t'-t) \Lambda^{-+}(t,t') + \theta(t-t') \Lambda^{+-}(t,t')$.
In all the calculations presented in this work we have introduced a small temperature of the order of $0.1\Gamma$ which
helps to stabilize the numerical calculations but does not produce significant deviations from the zero-temperature results.

%An special comment is needed for the evaluation of the self energy in time domain, which at zero temperature has a divergency. This divergency can be controlled by imposing a finite temperature and 
%performing a Matsubara expansion of the Fermi function. However, the evaluation of these kind of series are computationally demanding at low temperatures. In order to reduce the computational time, we 
%used the expansion proposed by T. Ozaki in reference \cite{Ozaki}, 
%where the series is performed using approximated Matsubara poles (purely imaginary poles with real residues)
%\begin{equation}
% f(w)=\sum_{n=-\infty}^{\infty}res_n\frac{e^{i\eta\beta_n}}{w-\beta_n}.Incluir??
%\end{equation}

\section{Single pole approximation}
\label{appendixC}

In this Appendix we discuss an approximation that can be used in order to compute the average current and population of the level
more efficiently. This approximation is useful in the regime
where the electron-phonon coupling is strong and the evaluation of the Fredholm determinant (\ref{z-nonint}) becomes computationally 
more demanding. For the evaluation of the current and dot 
population the  counting field is not needed and the 
transformation to the triangular representation in the Keldysh formalism can be performed
\cite{keldysh}. Then, the Dyson equation for the
retarded component of the Green function is given simply by 
\begin{eqnarray}
\label{Dyson_retarded}
G^{R}(t,t')=g^{R}_0(t,t')+\int_{0}^{t}{d t_2 \;K^{R}(t,t_2)G^{R}(t_2,t')}, \nonumber\\ 
\end{eqnarray}
where $K(t-t_2)=\int dt_1 g_0(t-t_1)\Sigma(t_1-t_2)$ is the Kernel of the integral equation. 
This kernel is time translational invariant, meaning that it only 
depends on the difference between time arguments, which implies that the retarded Green function, solution to the equation, 
preserves this symmetry. In this special case, Eq. (\ref{Dyson_retarded}) 
can be Laplace transformed arriving to
\begin{equation}
 G^{R}(s)=\frac{g^{R}_0(s)}{1-K^{R}(s)}.
\end{equation}
The issue of inverting the Laplace transform of a given function is in general not simple, because it usually involves the integration 
in a Bromwich contour of the function with a given collection of poles. However,
if $\Gamma\ll\omega_0$, the function has only a single dominant pole, $r$, and the retarded Green function of the dot 
can be written simply as
\begin{equation}
 \label{G^R}
 G^{R}(t-t')\approx \theta(t-t')\mbox{Res}(G^{R},s=r)\;e^{r(t-t')}\;,
\end{equation}
where Res indicates the residue at $s=r$.
This function is exponentially decaying, with a decay rate given by the real part of $r$.

For the calculation of the current and dot charge evolution we need 
the $G^{+-}$ component, which we analyze below for the initially empty case. For the initially occupied case, 
similar expressions can be derived by only 
changing the Green function component to the $G^{-+}$ one. The $G^{+-}$ component can be computed using the kinetic equation
\begin{equation}
 G^{+-}_{DTA}(t,t')=\int_0^t dt_1 dt_2 G^{R}(t,t_1)\Sigma^{+-}_{DTA}(t_1,t_2)G^{A}(t_2,t'),
\end{equation}
where the $\Sigma_{DTA}$ is the self-energy in our DTA approximation. The evolution of the average charge is 
given by the imaginary part of this Green function, i.e. $n_d(t)=\mbox{Im}[G^{+-}_{DTA}(t,t)]$.

Finally, when the coupling strength to both electrodes are equal ($\Gamma_L=\Gamma_R$), the symmetrized current can be computed as
\begin{eqnarray}
 \left\langle I(t)\right\rangle&=&\frac{\theta(t)}{2}\mbox{Re}\left\{\int_{0}^{t}dt_1\;G^{R}_{DTA}(t,t_1)\right.\nonumber\\
&& \times \left.\left[\Gamma_L f_L(t-t_1)-\Gamma_R f_R(t-t_1)\right]\right\},
\end{eqnarray}
where
\begin{eqnarray}
G^{R}_{DTA}(t,t_1)=\left[\Lambda^{+-}(t,t_1) G^{R}(t,t_1)\right.\nonumber\\
+\left.(\Lambda^{+-}(t,t_1)-\Lambda^{-+}(t,t_1))G^{+-}_{DTA}(t,t_1)\right].
\end{eqnarray}

\end{document}